\documentclass[twocolumn,usenatbib]{mn2e}

\usepackage{amsfonts}
\usepackage{amsmath}
\usepackage{amssymb}
\usepackage{aas_macros}
\usepackage{graphicx}
\usepackage{color}
\usepackage{hyperref}
\usepackage{float}

\def\lsim{~\rlap{$<$}{\lower 1.0ex\hbox{$\sim$}}}
\def\bsim{~\rlap{$>$}{\lower 1.0ex\hbox{$\sim$}}}

\def\hmsun{\ {\rm M_\odot/{\it h}}}

\def\ghz{\ {\rm GHz}}
\def\thz{\ {\rm THz}}

\def\la{\langle}

\def\dd{{\rm d}}
\def\ln{{\rm ln}}
\def\tr{{\rm tr}}

\def\mathbi#1{\textbf{\em #1}}

\def\kvh{\mathrm{\hat{\bf{k}}}}
\def\nvh{\mathrm{\hat{\bf{n}}}}

\def\vk{\mathbi{k}}
\def\vp{\mathbi{p}}

\def\vx{\mathbi{x}}

\def\vcc{\mathrm{C}}
\def\vff{\mathrm{F}}

\def\apjl{Astrophys.\ J.\ Lett.}

\def\aap{Astron.\ Astrophys.}

\def\apjs{Astrophys.\ J. Supp.}

\definecolor{RedWine}{rgb}{0.743,0,0}
\definecolor{RoyalBlue}{rgb}{0.25,.41,.88}
\definecolor{ForestGreen}{rgb}{.13,.54,.13}
\definecolor{DeepPurple}{rgb}{.72,.18,1}

\newcommand{\expf}[1]{{\rm e}^#1}  
\newcommand{\kB}{k_{\rm B}}  
\newcommand{\pot}[2]{{#1}\times 10^{#2}}  
\newcommand{\Planck}{{\it Planck}}  

\newcommand{\jens}[1]{{\color{blue} #1}}

\begin{document}

\title[Detecting the recombination signal]
{Detecting the cosmological recombination signal from space}

\author[Desjacques et al.]
{\parbox[t]{\textwidth}{Vincent Desjacques$^{1}$\thanks{Email: Vincent.Desjacques@unige.ch}, 
Jens Chluba$^{2,3}$\thanks{E-mail: jchluba@ast.cam.ac.edu}, Joseph Silk$^{3,4,5}$\thanks{Email: silk@iap.fr}, 
Francesco de Bernardis$^{6}$ \\ and Olivier Dor\'e$^{7,8}$}\\\,\\
$^1$ D\'epartement de Physique Th\'eorique and 
      Center for Astroparticle Physics (CAP), University of Geneva,  \\
      ~~~24 quai Ernest Ansermet, CH-1211 Geneva, Switzerland 
\\
$^2$ Kavli Institute for Cosmology Cambridge, Madingley Road, Cambridge, CB3 0HA, UK \\
$^3$ Department of Physics and Astronomy, Johns Hopkins University,  Baltimore MD 21218, USA \\
$^4$ CNRS \& UPMC, UMR 7095, Institut dÕAstrophysique de Paris, F-75014, Paris, France \\
$^5$ Laboratoire AIM-Paris-Saclay, CEA/DSM/IRFU, CNRS, Universite Paris Diderot,  F-91191 Gif-sur-Yvette, France \\
$^6$ Department of Physics, Cornell University, Ithaca, NY, USA 14853 \\
$^7$ Jet Propulsion Laboratory, California Institute of Technology, Pasadena, CA 91109 \\
$^8$ California Institute of Technology, Pasadena, CA 91125
}

\date{}
\label{firstpage}
\pagerange{\pageref{firstpage}--\pageref{lastpage}}

\maketitle

\begin{abstract}
Spectral distortions of the CMB have recently experienced an increased interest. One of the inevitable distortion signals of our cosmological concordance model is created by the cosmological recombination process, just a little before photons last scatter at redshift $z\simeq 1100$. These cosmological recombination lines, emitted by the hydrogen and helium plasma, should still be observable as tiny deviation from the CMB blackbody spectrum in the cm--dm spectral bands. In this paper, we present a forecast for the detectability of the recombination signal with future satellite experiments. We argue that  serious consideration for future CMB  experiments in space should be given to probing spectral distortions and, in particular, the recombination line signals. The cosmological recombination radiation not only allows determination of standard cosmological parameters, but also provides a direct observational confirmation for one of the key ingredients of our cosmological model: the cosmological recombination history. We show that, with present technology, such experiments are futuristic but feasible. The potential rewards won by opening this new window to the very early universe could be considerable.
\end{abstract}


\section{Introduction}
\label{sec:intro}
The cosmic microwave background (CMB) provides one of the cleanest sources of information about the Universe in which we live. In particular, the CMB temperature and polarisation anisotropies have allowed us to pin down the key cosmological parameters with unprecedented precision \citep{WMAP_params, Planck2013params}, and we are presently witnessing the final stages in the analysis of \Planck\ data \citep{Planck2015params}. Most of today's experimental effort is going into a detection of primordial $B$-modes, exploiting the curl polarisation patterns sourced by gravity waves created during inflation \citep{Kamionkowski1997, Seljak1997, Kamionkowski1998}. Many balloon or ground-based experiments, such as SPIDER, BICEP2, Keck Array, Simons Array, CLASS, etc. \citep[e.g.,][]{SPIDER, CLASS, BICEP2results, KeckArray2012}, are currently observing or coming online. These measurements will eventually exhaust all the information about the primordial Universe to be gained from the CMB anisotropies.

The next frontier in CMB measurements is the detection of spectral distortions \citep{Silk2014}. This is a field that has not changed since COBE/FIRAS set upper limits on the $\mu$ and $y$ parameters more than 20 years ago \citep{Fixsen1996}. In the standard cosmological model, chemical potential distortions \citep{Sunyaev1970mu} are expected at a level of about $10^{-4}$ of the FIRAS upper limits i.e., $\mu\simeq 10^{-8}$) as a consequence of the  damping of primordial adiabatic density fluctuations \citep{Sunyaev1970diss, Daly1991, Hu1994, Chluba2012}. A detection of this signal would probe the redshift range between thermalisation ($z\simeq \pot{2}{6}$) and recombination ($z\simeq 10^3$). While very interesting constraints could be derived for non-standard inflation scenarios \citep[e.g.,][]{Chluba2012inflaton}, since the distortion is cause by an integrated effect of energy injection in the early Universe, it would be challenging to distinguish a detection from more exotic sources of energy, such as damping of small-scale non-Gaussian density or blue-tilted primordial density fluctuations \citep{Hu1994isocurv, Dent2012, Chluba2013iso}, or even dark matter self-annihilation \citep{Chluba2013fore, Chluba2013PCA}.

A truly powerful probe of spectral distortions must carry redshift information. While the hybrid non-$\mu$/non-$y$ distortion \citep{Chluba2011therm, Khatri2012mix, Chluba2013Green} is limited to a narrow redshift range ($10^4 \lesssim z\lesssim \pot{3}{5}$), a far more powerful probe, capable of probing the Universe directly at unprecedentedly high redshift, would be the hydrogen and helium recombination line spectrum \citep{Dubrovich1975, Rybicki93, DubroVlad95, Dubrovich1997, Sunyaev2009}. This would be the jewel in the crown of any future spectral distortion experiment, capable of verifying the recombination history of the Universe and measuring the primordial helium fraction \citep[e.g.,][]{Chluba2008T0, Sunyaev2009}. It would simultaneously provide a unique calibration template for probing the origin of other signals such as an average $y$-distortion \citep{Zeldovich1969} caused by reionization and structure formation \citep{Hu1994pert, Cen1999, Refregier2000} or the aforementioned dissipation signal. A cosmic recombination line probe would require extensive frequency coverage over GHz to THz frequencies and exquisite sensitivity. It would open a new window on the Universe 380,000 years after the Big Bang  \citep{Sunyaev2009}. 

Here, we explore the feasibility of such a probe for a future space experiment. We will consider the following situation: low- and high-frequency channels are used to remove synchrotron radiation and thermal dust emission, respectively. At the remaining available frequencies, we can use a template for the recombination spectrum and cross-correlate it with the data. This matched-filtering approach is particularly well suited for our purpose as we are mainly interested in the detection level of the recombination spectrum. It will naturally take into account frequency correlations in the template. At the Fisher matrix level, this is equivalent to a forecast for the overall amplitude of the recombination spectrum template. 

The paper is organised as follows. We begin with a description of our model for the sky signal
in \S\ref{sec:model}, with  particular emphasis on the cosmic infrared background (CIB). 
We discuss experimental setups and present forecasts in \S\ref{sec:forecast}. We conclude in 
\S\ref{sec:conclusion}. We adopt a concordance, flat $\Lambda$CDM model in all illustrative calculations. 

\section{Modelling the residual intensity}
\label{sec:model}

We are interested in modelling the specific intensity (or brightness), $I_\text{res}(\nu)$, that 
remains after the subtraction of the average CMB blackbody spectrum as a function of frequency. 
Assuming that foregrounds such as dust etc. have been subtracted, this residual intensity is a 
sum of CMB spectral distortions, the CIB and noise induced by the incident radiation and the 
detector. We will now describe each of these components in more detail.

\subsection{Spectral distortions of the CMB}
\label{sub:CMB}

Upon subtracting the estimated CMB monopole, we are left with primordial $\mu$- and $y$-distortions, 
a temperature shift, which arises from our imperfect knowledge of the CMB temperature, and 
the cosmological recombination radiation. We shall neglect the residual (non-$\mu$, non-$y$) distortion 
signal related to the precise time-dependence of the energy release process \citep[e.g.,][]{Chluba2013Green}, 
which can add extra low intensity features to the broad primordial distortion. 

For the signals, we use the definitions commonly adopted in the treatment of CMB spectral 
distortions \citep[e.g.,][]{Chluba2011therm, Chluba2012}.
For the $\mu$-distortion, a small amount of energy $\Delta E$ is injected at constant photon 
number into a blackbody of reference temperature $T_0$. Once $\Delta E$ is fully comptonized,
we are left with a Bose-Einstein spectrum with chemical potential $\mu$ \citep{Sunyaev1970mu}. 
The intensity difference is 
\begin{align}
\label{eq:mu}
I^{\tiny \mu}\!(\nu) &=\mu\left(\frac{2h\nu^3}{c^2}\right)M(x) \\
& = \mu\left(\frac{2h\nu^3}{c^2}\right) G(x) \left(\frac{\pi^2}{18\zeta(3)}-\frac{1}{x}\right) 
\nonumber\;.
\end{align}
Here, $x=h\nu/\kB T_0$, $\kB$ and $h$ are the Boltzmann and Planck constants, and 
$G(x)=x\expf{x}/(\expf{x}-1)^2$ describes a pure temperature shift. Note also that 
$\pi^2/[18\zeta(3)]\approx 0.4561$.
The $y$-distortion is generated by inefficient diffusion of the photons in energy through 
scattering off of electrons \citep{Zeldovich1969}. The change in intensity reads
\begin{align}
\label{eq:y}
I^{\tiny y}\!(\nu) &= y \left(\frac{2h\nu^3}{c^2}\right) Y(x) \\
&= y\left(\frac{2h\nu^3}{c^2}\right)G(x) \left[x\left(\frac{\expf{x}+1}{\expf{x}-1}\right)-4\right]
\nonumber \;.
\end{align}
We must furthermore take into account the uncertainty in the temperature $T_0=2.726$K \citep{Fixsen1996, Fixsen2009} of the reference blackbody up to second order \citep{chluba/sunyaev:2004}.
At this order, the deviation from the reference blackbody intensity is the sum of a pure 
temperature shift and a $y$-distortions \citep{Chluba2013PCA},
\begin{equation}
\label{eq:t}
I^{\tiny T}\!(\nu) = \frac{2 h \nu^3}{c^2} 
\left[G(x) \Delta \left(1+\Delta\right)+Y(x)\frac{\Delta^2}{2} \right] \;.
\end{equation}
Hence, a relative error of $\Delta=5\times 10^{-4}\equiv \Delta T_0/T_0$ in the precise value 
of $T_0$ generates a $y$-distortion of amplitude $y\simeq 10^{-7}$. Patch-to-patch fluctuations 
in the CMB temperature will also contribute an average temperature shift \citep{chluba/sunyaev:2004}, 
which can simply be absorbed into the variable $\Delta$. 

The recombination spectrum, $I^\text{\tiny rec}(\nu)$, can be evaluated numerically with high 
accuracy. Since we are only interested in a detection of the recombination spectrum, we will
use the template obtained from the computation of\footnote{The data is available at:\\ \url{http://vivaldi.ll.iac.es/galeria/jalberto/recomb/}}
\citet{rubinomartin/chluba/sunyaev:2006,Chluba2006b, chluba/rubinomartin/sunyaev:2007} and \citet{rubinomartin/chluba/sunyaev:2008}. These calculations include the contributions from both hydrogen and helium. Refinements caused by helium feedback processes \citep{Chluba2009c} are omitted here, but should not affect the main conclusions of this work. Similarly, small changes in recombination radiation at low frequencies ($\nu\lesssim 1\ghz$) caused by recombinations to highly excited states \citep{Chluba2010, Yacine2013RecSpec} are omitted.

\subsection{Halo model of the CIB}
\label{sub:CIB}

Our model of the CIB is based on a halo model description of the relation between star-forming
galaxies and dark matter (sub)haloes \citep{shang/haiman/etal:2012}, which builds on the early 
works of \cite{knox/cooray/etal:2001} and \cite{amblard/cooray:2007}
\citep[see also][for a model based on conditional luminosity functions]{debernardis/cooray:2012}.

The average CIB brightness at a given frequency (in unit of Jy sr$^{-1}$) is
\begin{equation}
I^\text{\tiny CIB}(\nu)=\int_0^\infty\!\!\dd z\,\left(\frac{\dd\chi}{\dd z}\right)
a(z) \bar{j}_\nu(z) \;,
\end{equation}
where $\chi(z)$ is the line-of-sight comoving distance to redshift $z$ and
\begin{equation}
\bar{j}_\nu(z)=\int\!\!\dd L\,\bar{n}(L,z)\,\frac{L_{(1+z)\nu}}{4\pi}
\end{equation}
is the mean emissivity of galaxies below a certain flux limit at a frequency $\nu$ and per 
comoving volume . Here, $L_{(1+z)\nu}$ and $\bar{n}(L,z)$ denote the infrared galaxy luminosity 
(in W Hz$^{-1}$) and galaxy luminosity function, respectively, while $(1+z)\nu$ designates the 
rest-frame frequency. Following \cite{shang/haiman/etal:2012}, 
we split the mean emissivity into a sum of two contributions,
\begin{equation}
\bar{j}_\nu(z) = \int\!\!\dd M\,\frac{\dd N}{\dd M}(M,z)
\Bigl[f_\nu^\text{c}(M,z)+f_\nu^\text{s}(M,z)\Bigr] \;,
\end{equation}
where
\begin{align}
f_\nu^\text{c}(M,z) &= \frac{1}{4\pi}N_\text{c}L_{\text{c},(1+z)\nu}(M,z) \\
f_\nu^\text{s}(M,z) &= \frac{1}{4\pi} \int\!\!\dd m\,\frac{\dd n}{\dd m}(M,z) 
L_{\text{s},(1+z)\nu}(m,z)
\end{align}
are the average emissivity produced by the central and satellite galaxies of a given halo 
at redshift $z$, $\dd N/\dd M$ and $\dd n/\dd m$ are the halo and sub-halo mass functions,
and $M$ and $m$ are the parent halo and sub-halo masses. The numbers $N_{\rm c}$ and $N_{\rm s}$ of 
central and satellite galaxies are specified by the halo occupation distribution (HOD) 
\citep{berlind/weinberg:2002,zheng/berlind/etal:2005}. 
The total number of galaxies in a given halo of mass $M$ thus is $N_{\rm c}+N_{\rm s}$. 
Numerical simulations indicate that $N_{\rm c}$ typically follows a step-like function 
\citep{kravtsov/berlind/etal:2004}, while $N_{\rm s}$ can be parametrised by a power-law with 
logarithmic slope $\alpha\approx 1$. For the characteristic mass $M_\text{cen}$ of the 
step function, we ignore any luminosity dependence and adopt a value of $3\times 10^{11}\hmsun$ 
broadly consistent with the best-fit HOD models of \cite{zehavi/zheng/etal:2011}.
In all subsequent calculations, we use the halo and sub-halo mass functions provided by 
\cite{tinker/kravtsov/etal:2008} and \cite{tinker/wetzel:2010}, and integrate the sub-halo mass 
function from a minimum halo mass $M_\text{min}=10^{10}\hmsun$ to the parent halo mass $M$.

We assume that the same luminosity-mass relation holds for both central and satellite galaxies,
and relate the galaxy infrared luminosity to the host halo mass through the parametric relation
\citep{shang/haiman/etal:2012}
\begin{equation}
L_{(1+z)\nu}(M,z)=L_0\Phi(z)\Sigma(M)\Theta[(1+z)\nu] \;,
\end{equation}  
where $L_0$ is an overall normalisation that must be constrained from measurement of the CIB 
specific intensity in a given frequency range. The term $\Phi(z)$ describes the
redshift-dependence of the normalisation. We adopt the power-law scaling, 
$\Phi(z) = \left(1+z\right)^{3.6}$,
whereas, for the dependence $\Sigma(M)$ of the galaxy luminosity on halo mass, we assume a 
log-normal distribution with mean mass $M_\text{eff}=4.43\times 10^{12}\hmsun$ and variance 
$\sigma_\text{M/L}^2=0.5$ \citep[as in][]{planckcibpaper2014}. 
$M_\text{eff}$ characterises the peak and $\sigma_\text{M/L}$ the range of halo mass that 
produces a given luminosity $L$. Although our choice of $\sigma_\text{M/L}^2=0.5$ is somewhat
arbitrary, the CIB angular power spectra turn out to be fairly insensitive to 
$\sigma_\text{M/L}$ \citep{shang/haiman/etal:2012}. In addition, the value of $M_\text{eff}$
is consistent with the peak of the stellar-to-halo mass ratio inferred from semi-analytic
galaxy formation models \citep{guo/white/etal:2010}.

\begin{figure}
\center
\resizebox{0.48\textwidth}{!}{\includegraphics{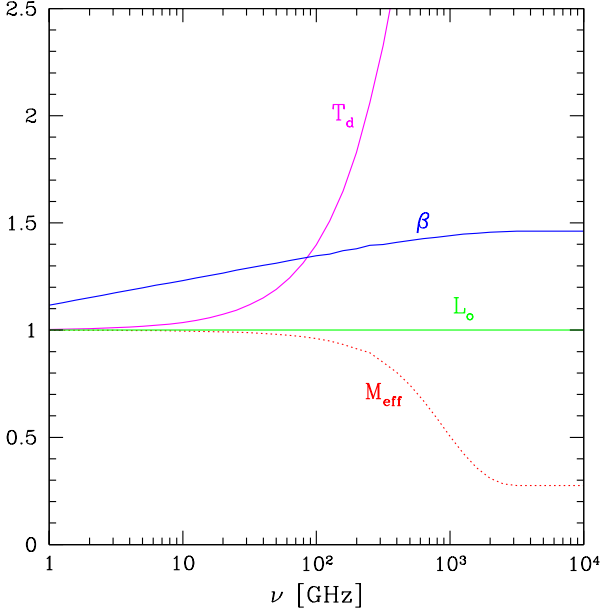}}
\caption{Logarithmic derivatives $\partial\ln I^\text{\tiny CIB}/\partial p_\alpha(\nu)$ 
of the average CIB intensity w.r.t. the model parameters $\vp=(M_\text{eff},T_d,\beta,L_0)$. 
The results are normalised such that all the derivatives are equal to unity for $\nu=0.1\ghz$.
Dotted curves indicate negative values.}
\label{fig:cib_derivs}
\end{figure}

For the galaxy spectral energy distribution (SED), we assume a modified blackbody shape with 
a power-law emissivity as in \cite{hall/keisler/etal:2010},
\begin{equation}
\Theta(\nu,z)\propto \left\lbrace\begin{array}{cc}
\nu^\beta B_\nu(T_d) & \nu\leq \nu_0 \\
\nu^{-\gamma} & \nu>\nu_0 \end{array}\right. \;,
\end{equation}
where $B_\nu(T)$ is the brightness of a blackbody with temperature $T$, and $T_d$ is the 
dust temperature. The grey-body and power-law connect smoothly at $\nu_0$ provided that
\begin{equation}
\frac{\dd\ln\Theta(\nu,z)}{\dd\ln\nu}\biggl\lvert_{\nu_0}=-\gamma \;. 
\end{equation}
Unlike \cite{serra/lagache/etal:2014}, however, we will treat both the dust emissivity index 
$\beta$ and the average dust temperature $T_d$ as free parameters, while we fix the SED index 
gamma to its mean value of $\gamma=1.7$ found by \cite{planckcibpaper2014}. The reason is
that, at frequencies $\nu\lesssim 10^3\ghz$, a change in the value of $\beta$ tilts the CIB
intensity in a way that could mimic a $\mu$/y-distortion or the broadband shape of the 
recombination spectrum.
This is illustrated in Fig.\ref{fig:cib_derivs}, in which we show logarithmic derivatives
of the average CIB intensity w.r.t. our free model parameters $(M_\text{eff},T_d,\beta,L_0)$
around the fiducial value 
$(M_\text{eff},T_d,\beta,L_0)=(4.43\times 10^{12}\hmsun,26$~K$,1.7,4.8\times 10^{-40}$ W Hz$^{-1}$).
Clearly, the logarithmic derivative $\partial\ln I^\text{\tiny CIB}/\partial\beta$ cannot be 
expressed as linear combination of the other derivatives, so we must treat $\beta$ as free
parameter since it is not well constrained by the data. We will adopt a fiducial value of 
$\beta=1.7$.

Note that the spectrum of star-forming galaxies typically features broad emission lines in 
mid-infrared (from 10 to 30$\mu$m) caused mainly by polycyclic aromatic hydrocarbon (PAH) 
molecules \citep[e.g.][]{leger/puget:1984,allamandola/tielens/etal:1985,smith/draine/etal:2007}, 
but these are far on the blue side of the spectrum ($\nu\gtrsim 10^4\ghz$).

\subsection{CIB anisotropies}
\label{sub:anisotropies}

While we will ignore fluctuations in the $\mu$ and $y$-distortions across the sky\footnote{The largest fluctuations are related to the spatially-varying $y$ distortion signal caused by warm-hot intergalactic medium and unresolved clusters \citep[e.g.,][]{Cen1999, Refregier2000, Miniati2000, Zhang2004}, while significant spatially-varying $\mu$ distortions can only be created by anisotropic energy release processes.}, we
must take into account anisotropies in the CMB and CIB intensity. In principle, it should be 
possible to subtract at least partly both CMB and CIB anisotropies from the observed patch using, e.g., data from the {\it Planck} experiment. In our fiducial analysis, however, we will assume they have not been removed. Therefore, we must include them in the Fisher information as they will contribute to the covariance of the signal, especially if the surveyed patch is small. 

The CIB intensity at a given frequency $\nu$ and in a given direction $\nvh$ can be expressed 
as the line of sight integral
\begin{equation}
\nonumber
I^\text{\tiny CIB}(\nu,\nvh)=
\int_0^\infty\!\!\dd z\,\left(\frac{\dd\chi}{\dd z}\right) a(z) \bar{j}_\nu(z) 
\left(1+\frac{\delta j_\nu(\chi(z)\nvh,z)}{\bar{j}_\nu(z)}\right).
\end{equation}
Using the Limber approximation \citep{limber:1954}, the angular cross-power spectrum of CIB 
anisotropies at observed frequencies $\nu$ and $\nu'$ is \citep{knox/cooray/etal:2001}
\begin{align}
\label{eq:ClCIB}
C_\ell^\text{\tiny CIB}\!(\nu,\nu') &= \int_0^\infty\!\!\frac{\dd z}{\chi^2}
\left(\frac{\dd\chi}{\dd z}\right) a^2(z)\, \bar{j}_\nu(z)\bar{j}_{\nu'}(z) \\
& \qquad \times P_j^{\nu\times\nu'}\!\!(k=\ell/\chi,z) \nonumber \;.
\end{align}
Assuming that spatial variations in the emissivity trace fluctuations in the 
galaxy number density, i.e. $\delta j/\bar{j}=\delta n_g/\bar{n}_g$, the power
spectrum $P_j$ is equal to the galaxy power spectrum $P_g$. In the framework of
the halo model discussed above, we thus have
\begin{equation}
P_j^{\nu\times\nu'}\!(k,z)= P_{1\text{h}}^{\nu\times\nu'}\!\!(k,z)
+P_{2\text{h}}^{\nu\times\nu'}\!\!(k,z)+ \text{shot-noise} \;.
\end{equation}
Hereafter, we will ignore the shot-noise contribution because it is typically smaller than the
2-halo term on large scales. 
The contributions from galaxy pairs in the same halo, the so-called 1-halo term, is
\begin{align}
P_{1\text{h}}^{\nu\times\nu'}\!\!(k,z) &= \frac{1}{\bar{j}_\nu\bar{j}_{\nu'}}
\int\!\!\dd M\,\frac{\dd N}{\dd M}\Bigl[f_\nu^\text{c}(M,z)f_{\nu'}^\text{s}(M,z)
u(k,z|M) \nonumber \\ & \qquad 
+ f_{\nu'}^\text{c}(M,z)f_\nu^\text{s}(M,z) u(k,z|M) \nonumber \\ & \qquad
+ f_\nu^\text{s}(M,z)f_{\nu'}^\text{s}(M,z) u^2(k,z|M) \Bigr] \;,
\end{align}
where $u(k,z|M)$ is the normalised Fourier transform of the halo density profile, 
whereas the 2-halo term reads
\begin{equation}
P_{2\text{h}}^{\nu\times\nu'}\!\!(k,z)=\frac{1}{\bar{j}_\nu\bar{j}_{\nu'}}
D_\nu(k,z) D_{\nu'}(k,z) P_\text{lin}(k,z) \;,
\end{equation}
with
\begin{align}
D_\nu(k,z) &= \int\!\!\dd M\,\frac{\dd N}{\dd M}\,b_1(M,z) u(k,z|M) \\
& \qquad \times \Bigl[f_\nu^\text{c}(M,z)+f_\nu^\text{s}(M,z)\Bigr] 
\nonumber \;.
\end{align}
Here, $P_\text{lin}$ is the linear mass power spectrum extrapolated to redshift $z$,
whereas $b_1$ is the (Eulerian) linear halo bias that follows from a peak-background
split applied to the halo mass function $\dd N/\dd M$. For consistency, we use the 
fitting formula given in \cite{tinker/robertson/etal:2010}.
The 1-halo term can only be seen with high angular resolution surveys reaching out to 
arcmin scales.

The CIB intensity reported in each pixel of the surveyed patch can be written as
\begin{equation}
I^\text{\tiny CIB}(\nu,\nvh_i) = \int\!\!\dd\nvh\,I^\text{\tiny CIB}(\nu,\nvh) B(\nvh,\nvh_i)
\end{equation}
where $B(\nvh,\nvh_i)$ is the beam profile. Specialising to azimuthally symmetric beam 
patterns, $B(\nvh,\nvh_i)$ is a function of $\nvh\cdot\nvh_i$ solely, and can thus be 
expanded in the Legendre polynomials $P_\ell$:
\begin{equation}
B(\nvh\cdot\nvh_i) = \frac{1}{4\pi}\sum_{\ell=0}^\infty \left(2\ell+1\right) B_\ell
P_\ell(\nvh\cdot\nvh_i) \;.
\end{equation}
While this is an approximation that is not met in most current experimental designs, 
it is straightforward to generalise the computation. Our main conclusions will not 
change. For a single Gaussian beam profile,
\begin{equation}
B(\nvh\cdot\nvh_i) = \frac{1}{2\pi\sigma_\text{B}^2}
\exp\left(-\frac{\vartheta^2}{2\sigma_\text{B}^2}\right) 
\end{equation}
with $\nvh\cdot\nvh_i=\cos\vartheta$ and $\sigma_\text{B}$ is related to the full width half 
maximum $\vartheta_\text{B}$ of the beam in radians through $\vartheta_B\approx 2.3548\sigma_\text{B}$. 
In this case, the series coefficients are well approximated by 
\citep{silk/wilson:1980,bond/efstathiou:1984,white/srednicki:1995}
\begin{equation}
\label{eq:beaml}
B_\ell \approx \exp\left(-\frac{1}{2}\ell(\ell+1)\sigma_\text{B}^2\right).
\end{equation}
The CIB intensity averaged over the surveyed patch reads
\begin{align}
\bar{I}^\text{\tiny CIB}(\nu) &= 
\frac{1}{\Omega_s}\int_{\Omega_s}\!\!\dd\nvh_i
\int\!\!\dd\nvh\,I^\text{\tiny CIB}(\nu,\nvh)B(\nvh\cdot\nvh_i) \\
&= \frac{1}{\Omega_s}\sum_{\ell m} B_\ell \int_{\Omega_s}\!\!\dd\nvh_i
\int\!\!\dd\nvh\,I^\text{\tiny CIB}(\nu,\nvh)
Y_\ell^{m\star}(\nvh) Y_\ell^m(\nvh_i) \nonumber \;.
\end{align}
Here, $\Omega_s=4\pi f_\text{sky}$ is the area of the surveyed patch. For an azimuthally symmetric 
patch centred at $\vartheta=0$ (for simplification), the integral over the unit vector $\nvh_i$ 
trivially is
\begin{equation}
\int_{\Omega_s}\!\!\dd\nvh_i Y_l^m(\nvh_i) \equiv W_l \delta_{m0}\jens{.}
\end{equation}
For a circular cap (i.e., unit weight for all pixels with polar angle such 
that $\vartheta\leq\vartheta_W $), the harmonic transform $W_\ell$ of the survey window is 
\citep[e.g,][]{manzotti/hu/benoitlevy:2014}
\begin{equation}
W_\ell = \sqrt{\frac{\pi}{2\ell+1}}\,\Bigl[P_{\ell+1}(x)-P_{\ell-1}(x)\Bigr]\nonumber \;.
\end{equation}
Here, $P_\ell(x)$ denote Legendre polynomials and $x=\cos\vartheta_W=1-2f_\text{sky}$ is the cosine 
of the opening angle. Substituting the multipole expansions for the survey mask and the beam profile, 
the partial sky average of $I^\text{\tiny CIB}(\nu,\nvh)$ simplifies to
\begin{equation}
\bar{I}(\nu) = \frac{1}{\Omega_s}
\sum_\ell W_\ell B_\ell \int\!\!\dd\nvh\,I^\text{\tiny CIB}\!(\nu,\nvh) Y_\ell^0(\nvh) \;.
\end{equation}
In the linear bias approximation, the Fourier modes of the CIB fluctuations are determined by
$\delta j_\nu(\vk,z)=\bar{j}_\nu(z) b(k,\nu,z)\delta(\vk,z)$. 
If we now successively write $\delta j_\nu(\chi(z)\nvh,z)$ as the Fourier transform of 
$\delta j_\nu(\vk,z)$, substitute the plane-wave expansion and use the Limber approximation, then the 
frequency-dependent multipoles, $a_{l0}(\nu)$, can eventually be written as the line of sight 
integral \citep{knox/cooray/etal:2001}
\begin{align}
\label{eq:al0}
a_{\ell 0}(\nu) &= \sqrt{4\pi} I^\text{\tiny CIB}\!(\nu) \delta_{\ell 0}+
i^\ell \frac{\sqrt{\pi}}{2}\int\!\!\frac{\dd^3k}{2\pi^2}\,
\frac{\Gamma\bigl((\ell+1)/2\bigr)}{\Gamma\bigl((\ell+2)/2\bigr)} \\
&\qquad \times a(\bar{z})D(\bar{z})\bar{j}_\nu(\bar{z})
\frac{b(k,\nu,\bar{z})}{k} \delta(\vk) Y_\ell^0(\kvh) \nonumber \;.
\end{align}
Here, $D(z)$ is the linear growth rate and  $\bar{z}$ is defined through the relation 
$k\chi(\bar{z})\approx \ell$. The monopole of the CIB intensity computed from patches of the sky 
with coverage fraction $f_{\rm sky}=\Omega_s/(4\pi)$ is thus given by
\begin{equation}
\bar{I}(\nu) = \frac{1}{\Omega_s}\sum_\ell W_\ell B_\ell\, a_{\ell 0}(\nu) \;,
\end{equation}
where the multipole coefficients $a_{l0}(\nu)$ are given by Eq.(\ref{eq:al0}). 
Of course, the average $\left\langle\bar{I}^\text{\tiny CIB}(\nu)\right\rangle$ over all such 
patches is exactly given by $W_0 B_0 a_{00}(\nu)\equiv I^\text{\tiny CIB}(\nu)$. However, owing to
cosmic variance, the monopole fluctuates from patch to patch with an amplitude given by
\begin{align}
\left[\delta\bar{I}^\text{\tiny CIB}(\nu)\right]^2 
&= \left\langle \bar{I}^\text{\tiny CIB}(\nu)^2\right\rangle
-\left\langle \bar{I}^\text{\tiny CIB}(\nu)\right\rangle^2 \\
&= \frac{1}{\Omega_s^2}\sum_{\ell,\ell'\geq 1} W_\ell W_{\ell'}^\star B_\ell B_{\ell'}^\star 
\bigl\langle a_{\ell 0}(\nu) a_{\ell'0}^\star(\nu)\bigr\rangle \nonumber \\
&= \frac{1}{\Omega_s^2}\sum_{\ell\geq 2}
\bigl\lvert W_\ell B_\ell\bigr\lvert^2 C_\ell^\text{\tiny CIB}\!(\nu,\nu) \nonumber \;,
\end{align}
where the cross-power spectrum $C_\ell^\text{\tiny CIB}\!(\nu,\nu)$ is given by Eq.(\ref{eq:ClCIB}). 
Note that, following \cite{knox/cooray/etal:2001}, we have used the Limber approximation and
approximated the ratio of Gamma functions squared in Eq.~\eqref{eq:al0} as $\approx 2/\ell$ to 
simplify the ensemble average $\bigl\langle a_{\ell 0}(\nu) a_{\ell'0}^\star(\nu)\bigr\rangle$.  
Therefore, our result strictly holds for $\ell\gg 1$. 

\begin{figure}
\center
\resizebox{0.48\textwidth}{!}{\includegraphics{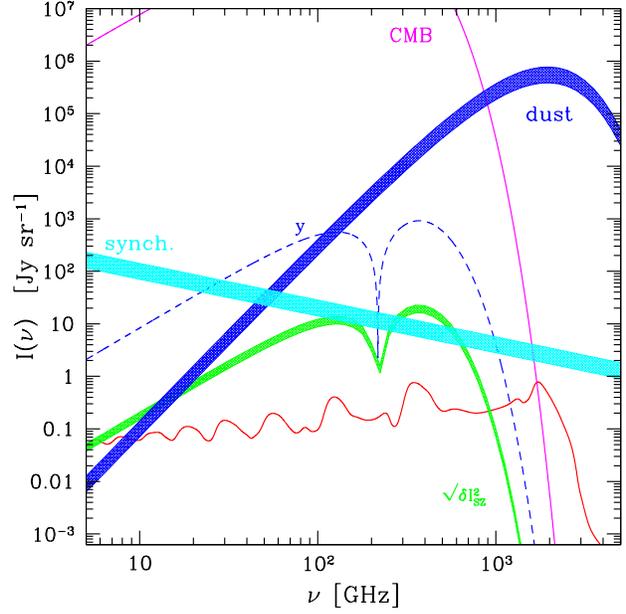}}
\caption{Contribution of galactic foregrounds and LSS-induced $y$-distortions to the 
intensity as a function of frequency. The various curves represent the CMB monopole, galactic dust 
and synchrotron emission. We have also shown the monopole of the $y$-distortion (primordial + LSS) 
with amplitude $y=5\times 10^{-7}$, as well as its expected rms variance (generated by LSS thermal
SZ effect) for a nearly all-sky survey ($f_\text{sky}=0.8$) with a beam FWHM of 1.6 deg. 
The specific intensity is in unit of Jansky per unit solid angle, Jy sr$^{-1}$, 
where 1 Jy = $10^{-26}$ W m$^{-2}$ Hz$^{-1}$.}
\label{fig:fgrnd1}
\end{figure}

\subsection{CMB primary and secondary anisotropies}

CMB anisotropies will also contribute to the covariance of the signal if they are not taken out. 
Ignoring the weak frequency dependence induced by Rayleigh scattering \citep[e.g.,][]{Lewis2013Ray}, 
the residual CMB intensity (i.e., with the reference blackbody spectrum subtracted) averaged over 
the surveyed patch reads 
\begin{equation}
\bar{I}^\text{\tiny CMB}(\nu)\approx I^{\tiny T}\!(\nu)\qquad \mbox{with}\qquad 
T \approx \Delta+ \bar{\Theta} \;.
\end{equation}
Here, $\Delta$ is the relative uncertainty on the temperature of the monopole and $\bar{\Theta}$ 
is the average temperature anisotropy in the surveyed window. In analogy with the CIB, the variance 
of CMB intensity fluctuations across different patches of the sky reads
\begin{equation}
\left[\delta\bar{I}^\text{\tiny CMB}(\nu)\right]^2 =\frac{1}{\Omega_s^2}\sum_{\ell\geq 2}
\bigl\lvert W_\ell B_\ell\bigr\lvert^2 C_\ell^\text{\tiny CMB}\!(\nu,\nu) \;,
\end{equation}
where the CMB intensity power spectrum is given by
\begin{equation}
C_\ell^\text{\tiny CMB}\!(\nu,\nu)\approx \left(\frac{2h\nu^3}{c^2}\right)^2 G^2(x)\, C_\ell \;.
\end{equation}
The contribution from Thomson scattering takes on the standard expression:
\begin{equation}
C_\ell=\frac{2}{\pi}\int_0^\infty\!\!dk\,k^2 g_{T\ell}^2(k) P_\Phi(k),
\end{equation}
where $g_{T\ell}(k)$ denotes the photon transfer function, which can be obtained from {\tt camb} 
\citep{CAMB}. We will assume that CMB intensity fluctuations are fully characterised by the 
primordial scalar amplitude $A_s$. 

The variance of primordial $\mu$- and $y$-distortions fluctuations (created before recombination) 
over patches can be ignored, since the current limits on the magnitude of $y$ and $\mu$ are fairly 
small so that fluctuations are expected to contribute at the $\delta I_\nu/I_\nu\simeq 10^{-10}$ level. 
Larger fluctuations of $\mu$ and $y$ across the sky could be created by the dissipation of acoustic 
modes with modulated small-scale power due to non-Gaussianity in the ultra-squeezed limit 
\citep{Pajer2012, Ganc2012}, however, current upper limits on $f_{\rm NL}\lesssim 2.5\pm 5.7$ 
\citep{Planck2015ng} suggest that this case is unlikely for scale-invariant non-Gaussianity, 
so that we ignore it here. 

Secondary $y$-distortions arise in the late-time Universe because CMB photons scatter off
hot electrons present in the gas of filaments and galaxy clusters \citep{SZeffect}. The magnitude 
of this $y$ distortion is an integral of the electron pressure along sight lines passing through 
the large scale structure (LSS). In a typical cluster, the electron temperature is $T_{\rm e}\sim 1$ 
keV and leads to $y\sim 10^{-6}$. Existing catalogues of galaxy clusters can be used to remove at 
least part of the signal generated at low redshift (i.e., large angular scales). Nevertheless, we 
will be left with a residual monopole, which can be absorbed into $I^y(\nu)$ as long as one is not 
interested in the primordial $y$-distortion, and a fluctuating contribution, whose angular power
spectrum is \citep[within the Limber approximation, see][]{persi/spergel/etal:1995,Refregier2000}
\begin{align}
C_\ell^\text{SZ}\!(\nu,\nu) &\sim y_0^2 \left(\frac{2h\nu^3}{c^2}\right)^2 Y^2(x) \\
&\quad \times \int_0^\infty\!\!\frac{\dd z}{\chi^2} \left(\frac{\dd\chi}{\dd z}\right)
\frac{\bar{T}_\rho^2}{a^4(z)}\,P_p\left(k=\ell/\chi,z\right)
\nonumber \;,
\end{align}
where the constant $y_0$ is $\sim 1.7\times 10^{-16}$ K$^{-1}$ Mpc$^{-1}$ for a helium fraction and 
baryon density consistent with Big-Bang Nucleosynthesis constraints, $\bar{T}_\rho$ is the 
volume-average, density-weighted gas temperature and $P_p(k,z)$ is the 3-dimensional power spectrum 
of pressure fluctuations. 

For simplicity, we will assume that $P_p(k,z)=b_p^2 P_\text{lin}(k,z)$ is 
a biased version of the linear mass power spectrum. We adopt $b_p=85$, which yields a prediction 
consistent with the simulations of \cite{Refregier2000} and, thus, provides a realistic upper limit 
to the signal.
We compute $\bar{T}_\rho$ as the halo virial temperature $T_\text{vir}$ weighted by the mass function, 
\begin{equation}
\bar{T}_\rho=\frac{1}{\bar{\rho}_{\rm m}}\int\!\!\dd M\,M\frac{\dd N}{\dd M}(M,z) T_\text{vir}(M,z)\;,
\end{equation}
where $\bar{\rho}_{\rm m}$ is the present-day average matter density, and
\begin{align}
T_\text{vir}(M,z) &= 6.03 \times 10^7 \beta^{-1} \left(\frac{\Delta_\text{vir}(z)}{18\pi^2}\right)^{1/3}
\left(\frac{M}{10^{15}\hmsun}\right)^{2/3} \nonumber \\
&\qquad \times \left(1+z\right) \Omega_{\rm m} \, {\rm K} \;.
\end{align}
Hence, $\bar{T}_\rho$ is a strongly decreasing function of redshift. 
We adopt $\Delta_\text{vir}\equiv 200$ at all 
redshift to be consistent with our choice of the halo mass function, and assume $\beta=2/3$ as in 
\cite{komatsu/kitayama:1999}. Furthermore, we only include halos in the mass range 
$M_\text{min}\leq M\leq M_\text{max}(z)$. While we consider a fixed minimum mass 
$M_\text{min}=10^{10}\hmsun$ ~\footnote{We have found that our predictions hardly change if we set 
$M_\text{min}=10^{12}\hmsun$.}, we allow $M_\text{max}(z)$ to vary with redshift to account for the 
possibility of removing the SZ contribution from low redshift clusters.

In Fig.~\ref{fig:fgrnd1}, we show the square-root of 
\begin{equation}
\left[\delta\bar{I}^\text{\tiny SZ}(\nu)\right]^2 =\frac{1}{\Omega_s^2}\sum_{\ell\geq 2}
\bigl\lvert W_\ell B_\ell\bigr\lvert^2 C_\ell^\text{\tiny SZ}\!(\nu,\nu)
\end{equation}
for an experiment with $f_\text{sky}=0.8$ and 1.6 deg angular resolution. The upper limit of the shaded
region assumes $M_\text{max}=10^{18}\hmsun$ at all redshifts. This corresponds to the signal that would 
be measured, had we not attempted to remove some of the $y$-distortion using external galaxy cluster 
catalogues. By contrast, the lower limit of the shaded region assumes that the SZ effect from groups and 
clusters with $T_\text{vir}\geq 1$ keV has been removed, so that $M_\text{max}(z)$ truly depends on 
redshift in our calculation. Note that these two limiting cases differ by $\lesssim$ 50\%, which shows 
that the contribution from virialized structures in filaments is quite substantial.
Overall, for the sky coverage adopted here, fluctuations in the $y$-distortions from the thermal SZ effect 
are of magnitude comparable to the recombination spectrum. 
Of course, $\left[\delta\bar{I}^\text{\tiny SZ}(\nu)\right]^2$ is exactly zero for an all-sky survey. 
For a ground-based survey targeting a 16 deg$^2$ patch of the sky as discussed below 
(see \S\ref{sub:experiments}), $\left[\delta\bar{I}^\text{\tiny SZ}(\nu)\right]^2$ is approximately ten 
times larger than shown in Fig.\ref{fig:fgrnd1}.

We will hereafter ignore the signal covariance induced by the LSS through the thermal SZ effect as we 
will either consider satellite missions with $f_\text{sky}\approx 1$, or a ground-based experiment for
which $\left[\delta\bar{I}^\text{\tiny CMB}(\nu)\right]^2$ and
$\left[\delta\bar{I}^\text{\tiny CIB}(\nu)\right]^2$ dominate the signal covariance. Notwithstanding, 
one should bear in mind that the LSS $y$-distortion could contribute significantly to the covariance
as well as bias the primordial $y$ interpretation if the sky fraction is significantly less than unity.

For comparison, we also show in Fig.\ref{fig:fgrnd1} the average $y$-distortion for $y=5\times 10^{-7}$, 
together with the approximate level of galactic emission from synchrotron and thermal dust (for which 
we assumed a power-law and grey-body spectrum, respectively). 
While we can reasonably handle the synchrotron emission with the low 
frequencies, removing the galactic dust monopole is more challenging as it is typically brighter than the 
CIB, and has a similar grey-body spectrum. Furthermore, the galactic dust emission fluctuates significantly 
at large angular scales ($\ell \lesssim 100$), with a power spectrum $C_\ell\propto \ell^{-2.6}$ (rather 
than $C_\ell\propto \ell$ for the CIB).

\section{Forecast for the recombination spectrum}
\label{sec:forecast}

Our model of the total (monopole) intensity $\bar{I}_\text{tot}(\nu)$ measured over a patch of 
the sky with coverage fraction $f_\text{sky}$ and in a frequency bin centred at $\nu_\alpha$ is
\begin{align}
\label{eq:model}
\bar{I}^\text{\tiny tot}\!(\nu_\alpha) &= 
I^{\tiny y}\!(\nu_\alpha) + I^{\tiny \mu}\!(\nu_\alpha) + I^{\tiny T}\!(\nu_\alpha) \\
& \qquad + I^\text{\tiny rec}\!(\nu_\alpha) 
+ \bar{I}^\text{\tiny CIB}\!(\nu_\alpha) + {\cal N}(\nu_\alpha)
\nonumber \;,
\end{align}
where $I^\text{\tiny rec}(\nu_\alpha)$ is the recombination signal which we wish to detect and the noise,
${\cal N}(\nu_\alpha)$, includes contribution from the incident radiation and from the detector. 
For simplicity, we have assumed that foreground emission within our own Galaxy from synchrotron radiation 
of cosmic ray electrons and thermal emission from dust grains has been separated out from the signal using 
low- and high-frequency channels. We leave a more detailed analysis of foreground removal (including
the CIB signal) for future work. 

\begin{figure}
\center
\resizebox{0.48\textwidth}{!}{\includegraphics{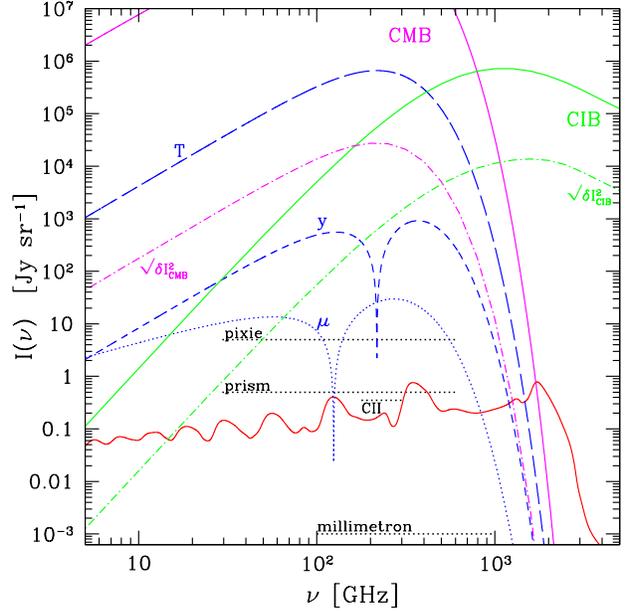}}
\caption{Average intensity as a function of frequency. The various curves represent the CMB and CIB 
monopole, the $T$, $\mu$ and $y$-distortions and the recombination spectrum. The dotted-dashed curves 
labelled as $\sqrt{\delta I_\text{\tiny CMB}^2}$ and $\sqrt{\delta I_\text{\tiny CIB}^2}$ indicate 
the r.m.s. variance of CMB and CIB fluctuations measured in a 16 deg$^2$ patch of the sky with a beam 
FWHM of 0.5 arcmin. The horizontal dotted lines show the frequency coverage and sensitivity expected for 
the {\small PIXIE}, {\small PRISM} and {\small MILLIMETRON} space experiments, as well as for a 
ground-based {\small CII} experiment. Note that the galactic dust monopole (not shown on this figure) 
is a grey-body similar to, though usually brighter than, the CIB (see Fig.~\ref{fig:fgrnd1}).}
\label{fig:fgrnd2}
\end{figure}

The various signal components are shown in Fig.~\ref{fig:fgrnd2}, together with
the frequency coverage and sensitivity expected for the {\small PIXIE} \citep{Kogut2011PIXIE}, 
{\small PRISM} \citep{PRISM2013WPII} and {\small MILLIMETRON} \citep{Smirnov2012} satellites, and a 
ground-based {\small CII} experiment \citep{gong/cooray/etal:2012}. 
The dotted-dashed curves represent the r.m.s. variance of CMB and CIB fluctuations in a patch of 16 deg$^2$ 
assuming a (Gaussian) beam FWHM of 0.5 arcmin. For such a small sky fraction 
($f_\text{sky}\sim 4\times 10^{-4}$), these are a few order of magnitudes larger than the signal in the 
frequency range 100--1000$\ghz$.

\subsection{Fisher matrix}
\label{sub:fisher}

Consider $N_\nu$ uncorrelated frequency bins centred at $\nu_\alpha$ and of bandwidth $\delta\nu$ 
covering the range $(\nu_\text{min},\nu_\text{max})$. Following \cite{tegmark/taylor/heavens:1997}, 
the Fisher matrix is given by\footnote{Although the trace is unnecessary here because the signal
is one-dimensional, we keep it for the sake of generality.}
\begin{equation}
\nonumber
\vff_{ij}=\sum_{\alpha=1}^{N_\nu}
\biggl[\frac{1}{2}\tr\!\left(\vcc_\alpha^{-1}\frac{\partial\vcc_\alpha}{\partial p_i}\vcc_\alpha^{-1}
\frac{\partial\vcc_\alpha}{\partial p_j}\right)
+\frac{\partial\langle{\vx}_\alpha\rangle}{\partial p_i}\vcc_\alpha^{-1}
\frac{\partial\langle{\vx}_\alpha\rangle}{\partial p_j}\biggr] \;,
\end{equation}
where $\langle{\vx}_\alpha\rangle=\bigl\langle \bar{I}^\text{\tiny tot}\!(\nu_\alpha)\bigr\rangle$ 
is the ensemble average of the residual intensity over random realisations of the surveyed patch, 
$\vcc_\alpha$ is the covariance of $I^\text{\tiny tot}\!(\nu_\alpha)$, and $\vp$ is the vector of 
model parameters. In our model, the signal ensemble average is
\begin{equation}
\langle{\vx}_\alpha\rangle = I^{\tiny y}\!(\nu_\alpha) + I^{\tiny \mu}\!(\nu_\alpha)
+ I^\text{\tiny rec}\!(\nu_\alpha) + I^{\tiny T}\!(\nu_\alpha) 
+ \bar{I}^\text{\tiny CIB}\!(\nu_\alpha) \;,
\end{equation}
whereas the covariance of the data is given by
\begin{align}
\vcc_\alpha &=
\left[\delta\bar{I}^\text{\tiny CIB}(\nu_\alpha)\right]^2
+\left[\delta\bar{I}^\text{\tiny CMB}(\nu_\alpha)\right]^2
+\bigl\langle{\cal N}^2(\nu_\alpha)\bigr\rangle \\
&= \frac{1}{\Omega_s^2}\sum_{\ell\geq 2} \bigl\lvert W_\ell B_\ell\bigr\lvert^2 
\Bigl[C_\ell^\text{\tiny CIB}\!(\nu_\alpha,\nu_\alpha)
+C_\ell^\text{\tiny CMB}\!(\nu_\alpha,\nu_\alpha)\Bigr] +\delta I_\nu^2 
\nonumber \;.
\end{align}
For an all-sky survey, the covariance is simply $C_\alpha=\delta I_\nu^2$. 
Furthermore, we ignore dependence on cosmological parameters and assume that the recombination signal 
is perfectly known, so that we can perform an idealised 
cross-correlation analysis. The model parameters therefore are 
$\vp=(\mu,y,\Delta,A_s,M_\text{eff},T_d,\beta,L_0,A_r)$, 
where $A_r$ is the amplitude of the recombination signal which we seek to constrain. Our fiducial 
parameter values are $\mu=10^{-7}$, $y=5\times 10^{-7}$, $\Delta=5\times 10^{-4}$ and $A_s=2.2\times 10^{-9}$,
together with CIB parameters $M_\text{eff}=4.43\times 10^{12}\hmsun$, $T_d=26$ K, $\beta=1.7$ and 
$L_0=4.8\times 10^{-40}$ W Hz$^{-1}$.

\begin{table*}
\caption{Experimental setups considered for the Fisher matrix forecast. The sensitivity of the ground-based 
experiment was derived for a 10m aperture telescope at frequency 238$\ghz$ (see text for details).}
\vspace{1mm}
\begin{center}
\begin{tabular}{cccccc} 
\hline
 & $\Omega_s$ & $\vartheta_\text{B}$ & $\delta I_\nu$ [Jy sr$^{-1}$] & $\delta\nu$ [GHz] & $\nu$-coverage [GHz] \\
\hline\hline
{\small PIXIE} & full sky & 1.6 deg & 5 & 15 & 30 -- 600 \\
\hline
{\small PRISM} & full sky  & 1.6 deg & 0.5 & 15 & 30 -- 600 \\
\hline
{\small MILLIMETRON} & full sky  & 3 arcmin & $10^{-3}$ & 1 & 100 -- 1000 \\
\hline
{\small GROUND CII} & 16 deg$^2$  & 0.5 arcmin & 0.35 & 0.4 & 185 -- 310 \\
\hline
\end{tabular}
\end{center}
\label{table1}
\end{table*}

\subsection{Experimental setup}
\label{sub:experiments}

The sensitivity of an ideal instrument is fundamentally limited by the noise of incident photons. 
This is already the case of the best bolometric detectors. For photon noise-limited detectors, 
a gain in sensitivity can only be achieved by collecting more photons through an increase in the 
number of detectors, collecting area and/or integration time.

For the moment, we assume ${\cal N}(\nu)$ does not exhibit frequency correlations
and is normally distributed with a variance $\bigl\langle{\cal N}^2\bigr\rangle=\delta I_\nu^2$
which depends only on the bandwidth $\delta\nu$. The analysis is performed assuming top-hat 
frequency filters, so that the filter response is 1 if $\nu\in[\nu_\alpha-\delta\nu/2,\nu_\alpha+\delta\nu/2]$ 
and zero otherwise. Regarding the frequency coverage, it is difficult to sample frequencies less 
than $\sim 30\ghz$ from space, for which the wavelength becomes larger than the typical size of 
the device (the collecting area) developed to measure them. However, this frequency regime can be 
targeted from  the ground \citep[see][for a recent analysis]{rao/subrahmanyan/etal:2015}.
Finally, the spectral resolution will be $\delta\nu=15\ghz$ for the {\small PIXIE} experiment, 
but it could be as high as $1\ghz$ for a satellite mission like {\small MILLIMETRON}. 
For the sake of comparison, we will also consider a ground-based {\small CII} mapping instrument which 
achieve sub-Gigahertz resolution and targets $\mathcal{O}(100)\ghz$ frequencies corresponding to 
{\small CII} fine structure lines. We refer the reader to \cite{rao/subrahmanyan/etal:2015} for 
ground-based surveys probing the $\lesssim 1\ghz$ frequency range.
Table \ref{table1} summarises the sensitivities of the various setups we consider in our Fisher
forecast. Note that the lowest frequency measured is $30\ghz$.
 
\begin{figure}
\center
\resizebox{0.48\textwidth}{!}{\includegraphics{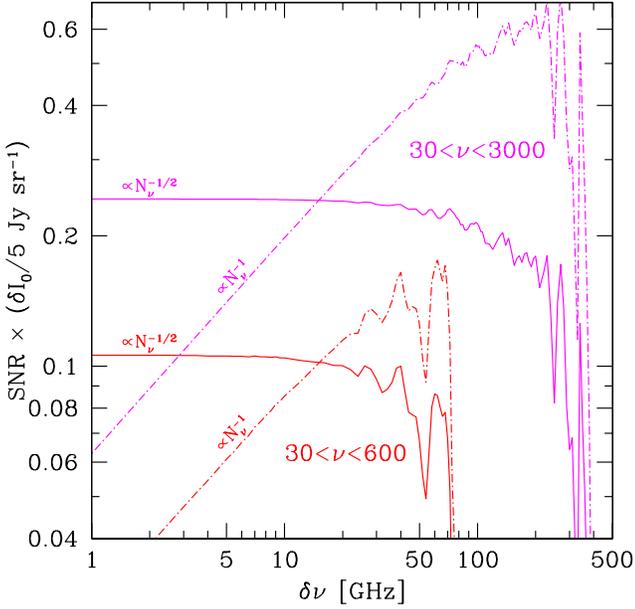}}
\caption{Signal-to-noise ratio for the detection of the recombination spectrum as a function
of spectral resolution after marginalisation over the remaining model parameters. 
We consider two different frequency coverage, $30<\nu<600$ and 
$30<\nu<3000\ghz$; as well as two different scaling with the number of frequency channels: 
$\propto N_\nu^{-1/2}$ (standard square-root law detectors) and $\propto N_\nu^{-1}$
(Fourier transform detectors). The SNR was normalised such that the two different types of
detectors yield the same SNR for $\delta\nu=15\ghz$.}
\label{fig:snr}
\end{figure}

We estimate the sensitivity of the ground-based experiment from the requirements given in 
\cite{gong/cooray/etal:2012}. Specifically, we consider a single-dish experiment with aperture
diameter $D=10$m. At $238\ghz$ \citep[which corresponds to CII emission line at redshift $z=7$ 
through the transition $^2P_{3/2}\to 2P_{1/2}$, see][]{gong/cooray/etal:2012}, the resulting 
beam FWHM is approximately $\vartheta_\text{B}=1.22\lambda/D\approx 0.53$ arcmin. The equivalent 
beam area is $\Omega_\text{B}\approx 8.82\times 10^{-5}$ deg$^2$. Assuming a noise equivalent 
flux density of NEFD=10 mJy s$^{1/2}$ per spectral resolution element $\delta\nu=0.4\ghz$ 
\citep[consistent with the parameters in][]{gong/cooray/etal:2012}, we can estimate the 
sensitivity from the relation
\begin{equation}
\delta I_\nu = \frac{{\rm NEFD}}{\sqrt{2\tau}\Omega_\text{B}}\;,
\end{equation}
where $\tau$ is the total integration time and the factor of $\sqrt{2}$ arises from the 
Nyquist sampling and from the assumption of optical chopping\footnote{The source is measured one-half 
of the time, and one must differentiate the input signals.}.  For a total integration time of
4000 hours, we find $\delta I_\nu=50$ Jy sr$^{-1}$ for a single detector. Averaging over
20000 bolometers, we eventually obtain $\delta I_\nu=0.35$ Jy sr$^{-1}$.

Let us assume that the frequency range is evenly split into frequency channels of spectral
resolution $\delta\nu_0$, and let $\delta I_0$ be the channel sensitivity that can be achieved 
at this resolution. Since the recombination spectrum features emission lines with a 
characteristic width of the order of their peak frequency (see Fig.~\ref{fig:fgrnd2}), we 
expect that the signal-to-noise ratio (SNR) for a detection of the recombination spectrum saturates 
below a certain spectral resolution. This, however, will be true only if the detector follows
the standard square-root law SNR $\propto N_\nu^{-1/2}$, where $N_\nu$ is the number of frequency
channels corresponding to a spectral resolution $\delta\nu$. For a Fourier transform spectrometer 
(FTS), as utilised in PIXIE and PRISM, the scaling turns out to be 
SNR $\propto N_\nu^{-1}$. 

To see this, we follow \cite{Kogut2011PIXIE} and consider a total integration time $\tau$. 
If $N_s$ is the number of time-ordered samples used for the Fourier transform, then each sample is 
observed during a time $\tau/N_s$. Therefore, the noise in each time-ordered sample of the sky 
signal $S(t_i)$ is
\begin{equation}
\delta S(t_i) = \frac{\mbox{NEP}}{\sqrt{\tau/2N_s}} \;,
\end{equation}
where the factor of two converts between time and frequency domains (the number of frequency
channels is $N_\nu=N_s/2$) and NEP is the noise 
equivalent power of the detected radiation. Note that NEP generally is a function of frequency. 
If the noise from different time measurement $t_i$ is uncorrelated, then in each frequency bin of 
the FTS it is
\begin{equation}
\delta S(\nu_\alpha)=\frac{\delta S(t_i)}{\sqrt{N_\nu}} = \frac{2 \mbox{NEP}}{\sqrt{\tau}}\;,
\end{equation}
independent of the spectral resolution $\delta\nu$ (at fixed $N_\nu\delta\nu$). Therefore, 
since the sensitivity is $\delta I_\nu(\nu_\alpha) \propto \delta S(\nu_\alpha)/\delta\nu$,
we obtain the scaling $\delta I_\nu\propto \delta\nu^{-1}\propto N_\nu$. This implies that the 
signal-to-noise behaves like SNR $\propto \delta I_\nu^{-1}\propto N_\nu^{-1}$. 
This also means that (even under idealised conditions) averaging the SNR of individual synthesised 
FTS channels (improvement of $\simeq \sqrt{N_\nu}$) does not regain you the same sensitivity as 
directly measuring at lower frequency resolution.

In the FTS case, we thus expect the SNR to reach a maximum at some optimal spectral resolution 
before falling again as $\delta\nu$ further decreases. 
Namely, under a change $\delta\nu_0\to\delta\nu$ in spectral resolution, the sensitivity of the 
detector becomes
\begin{equation}
\delta I_\nu=\delta I_0 \left(\frac{\delta\nu_0}{\delta\nu}\right)^\beta = \delta I_0
\left(\frac{N_\nu}{N_0}\right)^\beta \;,
\end{equation}
where $N_\nu\delta_\nu=N_0\delta\nu_0$, while the signal-to-noise for, say, the model parameter 
$p_i$ scales like
\begin{equation}
(\text{SNR})^2=\frac{1}{\delta I_0^2}\left(\frac{N_0}{N_\nu}\right)^{\!2\beta}
\sum_{\alpha=1}^{N_\nu} \left(\frac{\partial I}{\partial p_i}(\nu_\alpha)\right)^2 \;.
\end{equation}
Here, $\beta=1/2$ and $1$ for a square-root law and Fourier transform detector, respectively.
Furthermore, we have assumed that both detectors yield the same SNR at the spectral resolution
$\delta\nu_0$.

\begin{figure}
\center
\resizebox{0.48\textwidth}{!}{\includegraphics{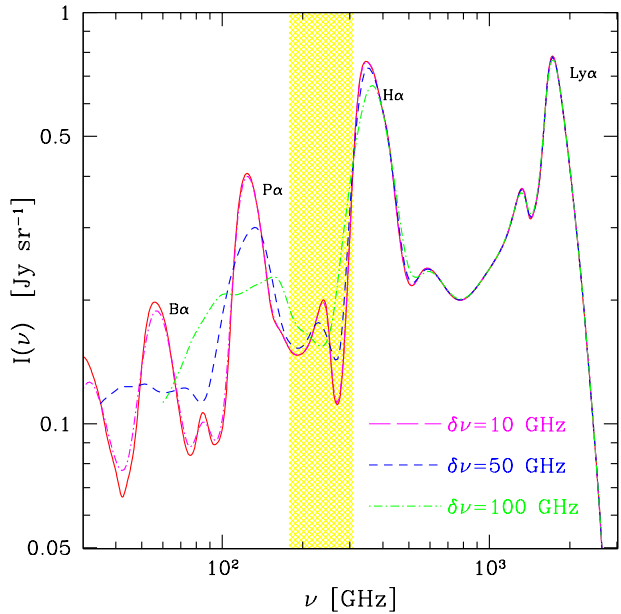}}
\caption{Effect of spectral resolution on the recombination spectrum. At spectral resolution
$\delta\nu\gtrsim 100\ghz$, the Paschen-$\alpha$ line is wiped out and, unless the frequency
coverage is wide enough to include the Lyman-$\alpha$ line, the SNR drops sharply (see Fig.
\ref{fig:snr}). The shaded region indicates the frequency range probed by the ground {\small CII}
experiment. In this spectral region, the narrow feature at $\nu\approx 270\ghz$ is the combination
of two lines with positive and negative peak intensity imprinted during {\small HeII}$\to${\small HeI} 
recombination \citep[see][]{rubinomartin/chluba/sunyaev:2008}.} 
\label{fig:rec_smooth}
\end{figure}

Figure \ref{fig:snr} shows the SNR of the recombination spectrum as a function of spectral 
resolution for an all-sky satellite experiment that measures the intensity monopole, 
Eq.~\eqref{eq:model}, with a sensitivity $\delta I_\nu=5\,$Jy sr$^{-1}$ at spectral resolution 
$\delta\nu=15\ghz$, very close to the {\small PIXIE} specifications.
We consider two different frequency ranges extending up either to $600\ghz$ or $3\thz$, so that the 
Lyman-$\alpha$ (Ly$\alpha$) line is only included in the latter case. The other components of 
our model [see Eq.~\eqref{eq:model}] have all been marginalised over.
Consider a standard square-root detector and measurements in the range $30<\nu<600\ghz$ for
instance. In this case, the SNR decreases mildly as the spectral resolution increases from 10 to 
$120\ghz$, and the Brackett-$\alpha$ (B$\alpha$) line is gradually smoothed out. 
The SNR drops abruptly around $\delta\nu\gtrsim 120\ghz$, which corresponds to the disappearance
of the Paschen-$\alpha$ (P$\alpha$) line. Therefore, one is basically left with only one line,
the Balmer-$\alpha$ (H$\alpha$), at spectral resolution $\delta\nu\gtrsim 120\ghz$ unless one 
extends the measurements out to $\simeq 3\thz$ to include the Lyman-$\alpha$ line 
(this is quite apparent in Fig.~\ref{fig:rec_smooth}). In both cases, the SNR 
saturates for $\delta\nu\lesssim 10\ghz$ for a standard square-root detector whereas, for a FTS, 
the largest SNR is obtained for $\delta\nu\sim 50$ and $\sim 200\ghz$. Note that the spectral 
resolution of {\small PIXIE} is $15\ghz$, which is close to optimal for a standard square-root 
detector when the Lyman-$\alpha$ line is not measured.

\subsection{Mitigating the CIB contamination}

In order to assess the extent to which the CIB degrades the SNR of the recombination spectrum, 
consider a {\small PIXIE} experiment with sensitivity $\delta I_\nu=5\,$Jy sr$^{-1}$, spectral
resolution $\delta\nu=15\ghz$ and frequency range $30<\nu<600\ghz$. Assuming we have perfect 
knowledge of all the model parameters except the amplitude of the recombination signal, the SNR 
for the recombination spectrum is SNR $\simeq$ 0.44. On marginalising over the four parameters 
$(M_\text{eff},T_d,\beta,L_0)$ describing the CIB, the SNR drops to $0.17$. 
Further marginalisation over the primordial CMB spectral distortions $y$ and $\mu$ brings the 
SNR down to 0.10. For the
wider frequency coverage $30<\nu<3000\ghz$, the degradation in the SNR following marginalisation
over the CIB is nearly a factor of 4 (down from 0.96 to 0.26).
Unsurprisingly, the amplitude of the recombination spectrum $A_r$ strongly correlates with 
$M_\text{eff}$, $\beta$ and $L_0$ (with a correlation coefficient $|r|\approx 0.9$ in all cases).
This is another justification for treating $\beta$ as free parameter. 
This also highlights that the continuum part of the cosmological recombination radiation is more 
difficult to isolate, even if its amplitude may be close to the sensitivity. Accessing the variable 
component with its quasi-periodic features strongly improves the situation.
For an experiment like {\small PRISM}, the recombination spectrum could be detected with a 
signal-to-noise ratio of SNR$\,\simeq 2.4$ if the frequency coverage extends up to $3\thz$. 
Finally, for the {\small MILLIMETRON} experiment, the SNR is as large as $\simeq 2500$ for the 
parameter values quoted in Table \ref{table1}.

\begin{figure}
\center
\resizebox{0.48\textwidth}{!}{\includegraphics{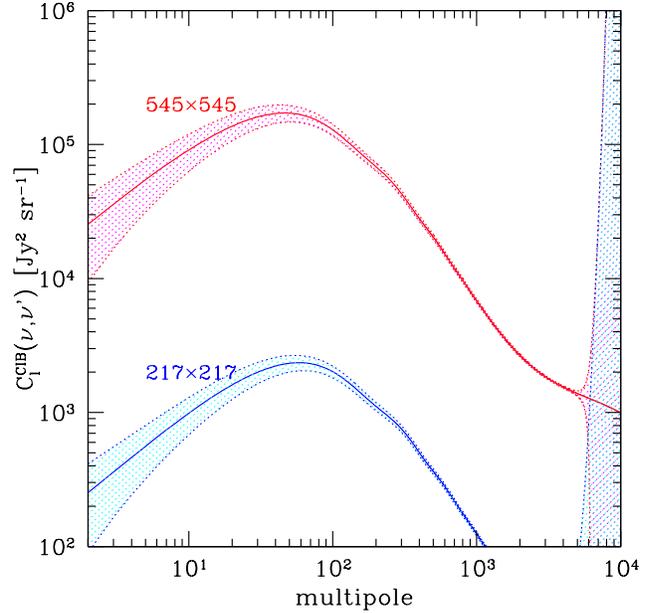}}
\caption{CIB angular power spectrum as measured by an experiment with {\small PRISM} sensitivity
and a beam angular resolution of 3 arcmin. For the spectral resolution considered, $38\times 38$ 
cross-power spectra $C_\ell^{\nu\nu'}$ can be computed at each multipole. 
Notice the flattening at $\ell\gtrsim 1000$ caused by the 1-halo term.}
\label{fig:cib_cl}
\end{figure}

To ascertain the extent to which our CIB model parameters could be constrained by a measurement
of the CIB anisotropies, we consider the Fisher matrix \citep[e.g.,][]{penin/dore/etal:2012}
\begin{equation}
\label{eq:fishcib1}
F_{ij}^\text{\tiny CIB}=\frac{f_\text{sky}}{2}\sum_\ell (2\ell+1)
{\rm tr}\!\left[{\bf C}_\ell^{-1}\frac{\partial{\bf C}_\ell}{\partial p_i}{\bf C}_\ell^{-1}
\frac{\partial{\bf C}_\ell}{\partial p_j}\right]\;.
\end{equation}
Here, ${\bf C}_\ell$ is the full, $N_\nu\times N_\nu$ covariance matrix at a given multipole 
$\ell$. The multiplicative factor of $f_\text{sky}$ reflects the fact that, for partial
sky coverage, modes with a given multipole are partially correlated, hence the variance of
each measured $C_\ell$ is larger. The matrix ${\bf C}_\ell$ takes the form
\begin{equation}
{\bf C}_\ell = \left(\begin{array}{ccc}
C_\ell^{\nu_1\nu_1}+{\cal N}_\ell^{\nu_1} & C_\ell^{\nu_1\nu_2} &  \\
C_\ell^{\nu_2\nu_1} & C_\ell^{\nu_2\nu_2}+{\cal N}_\ell^{\nu_2} & \\
& & \ddots
\end{array}\right) \;,
\end{equation}
where, for shorthand convenience, 
$C_\ell^{\nu_i\nu_j}\equiv C_\ell^\text{\tiny CIB}(\nu_i,\nu_j)$ and 
${\cal N}_\ell^\nu\equiv {\cal N}_\ell(\nu)$ is the angular power spectrum of the instrumental 
noise. Assuming uncorrelated and isotropic pixel noise, the latter is given by 
$\Omega_\text{\tiny pix}\sigma_\text{\tiny pix}^2 B_\ell^{-2}
\equiv f_\text{sky}\omega^{-1}B_\ell^{-2}$ \citep{knox:1995}, where 
$\Omega_\text{\tiny pix}$ is the solid angle subtended by one pixel, $\sigma_\text{\tiny pix}$ 
is the noise per pixel and $B_\ell$ is the experimental beam profile, Eq.~\eqref{eq:beaml}. 
Since our convention is to express the brightness $I(\nu)$ in unit of Jy sr$^{-1}$, the CIB 
angular power spectra $C_\ell^{\nu_1\nu_2}$ are in unit of Jy$^2$ sr$^{-1}$. 
The noise per pixel squared is $N_\text{\tiny pix}\delta I_\nu^2$ whereas the pixel solid angle is
$4\pi/N_\text{\tiny pix}$.
Therefore, the weight $\omega^{-1}$ is simply $\omega^{-1}=4\pi\delta I_\nu^2$ and, as expected, 
is independent of the sky pixelization. 
For the {\small PIXIE}- and {\small PRISM}-like specifications, we find $\omega^{-1}=314$ and 
3.14 Jy$^2$ sr$^{-1}$, respectively.
For illustration, Fig.~\ref{fig:cib_cl} shows the CIB angular power spectrum at two different 
frequencies for a {\small PRISM}-like experiment with 3 arcmin angular resolution of the imager.  

We have evaluated the Fisher matrix Eq.(\ref{eq:fishcib1}) for these experimental setups 
with a conservative value of $f_\text{sky}=0.8$, and subsequently computed the SNR for the 
recombination spectrum upon adding the Fisher information, $F_{ij}=F_{ij}^\text{mon}+F_{ij}^\text{cib}$,
where $F_{ij}^\text{mon}$ is the Fisher matrix for a measurement of the intensity monopole. 
For our fiducial beam FWHM of 1.6 deg (with $l_\text{max}=1000$), the improvement is negligible. 
A measurement of the CIB anisotropies at a much finer angular resolution of 3 arcmin (for which 
we adopt $l_\text{max}=10^4$) somewhat increases the SNR, but the improvement remains small.
Namely, for the {\small PIXIE} and {\small PRISM}-like experiments respectively, the SNR 
for a measurement of the recombination spectrum increases from 0.10 to 0.16 and from 1.02 to 1.15 
when the angular resolution is changed from 1.6 deg to 3 arcmin. This highlights that, in terms of 
the recombination signal, increased angular resolution is not a main issue. 
We have not pushed the angular resolution below 3 arcmin as our model does not include shot noise, 
which we expect to contribute significantly to the $C_\ell^{\nu\nu}$ at multipoles $\ell\gtrsim 10^4$.

We have not computed $F_{ij}^\text{cib}$ for the small field {\small CII} survey because the 
numerical evaluation is very time-consuming, owing to the exquisite frequency resolution. 
Nevertheless, Fig.~\ref{fig:fgrnd2} clearly shows that, since patch-to-patch fluctuations in 
the CIB are a few orders of magnitude larger than the recombination signal at frequencies 
$\sim 200\ghz$, removing the CIB is an even greater challenge for ground-based surveys targeting 
those frequencies. In the idealised situation where the CIB is perfectly known and the patch has
been cleaned from CMB anisotropies, the signal-to-noise ratio is SNR~$\simeq 2.74$. The feature 
at $\nu\approx 270\ghz$ (see Fig.~\ref{fig:rec_smooth}) helps distinguish the recombination 
spectrum from the CMB spectral distortions. Overall, our Fisher 
analysis shows that an all-sky survey covering a large frequency range at moderate frequency 
resolution $\Delta\nu\sim 10\ghz$ will perform much better than an experiment targeting a small 
patch of the sky at high spectral resolution ($\Delta\nu\sim 0.1\ghz$). 

\section{Conclusions}
\label{sec:conclusion}

We have presented a Fisher matrix forecast for the detection of the recombination line 
spectrum with future space experiments. The main caveat of our analysis is the assumption 
that galactic synchrotron and dust emission have been separated out using low- and 
high-frequency channels. We have also assumed that most of the thermal SZ effect generated 
by the large-scale structure can be removed using external catalogues of galaxy clusters
when it contributes significantly to the signal covariance (when $f_\text{sky}\ll 1$). 
Aside from that, our main findings can be summarised as follows:

\begin{itemize}
\item The cosmic infrared background is the main contaminant to the recombination signal. 
Detecting the recombination lines requires sub-percent measurement of the CIB spectral 
shape at frequencies $\nu\lesssim 1\thz$. The $T$, $y$ and $\mu$-distortions only weakly
correlate with the recombination spectrum if the frequency coverage is large enough.
\item Adding information from the CIB angular power spectrum does not greatly improve the SNR, 
even at arcmin angular resolution. Note that we have not considered the possibility of 
cross-correlating the CIB with  large-scale structure to better constrain the CIB model 
parameters.
\item While a ground-based {\small CII}-like experiment targeting a small patch of the sky 
cannot beat the CIB fluctuations, even with very high spectral resolution and exquisite 
sensitivity, an all-sky satellite mission can do this.
\item For an all-sky measurement in the frequency range $30\leq\nu\leq 600\ghz$, a spectral 
resolution $\delta\nu\lesssim 10\ghz$ is optimal at fixed $\delta\nu\delta I_\nu^2$. For a 
Fourier transform spectrometer, the optimal spectral resolution is larger 
($\delta\nu\simeq 50\ghz$) and dependent on the frequency coverage (see Fig.~\ref{fig:snr}), 
owing to the noise-scaling $\delta I_\nu \propto \delta\nu^{-1}$.
\item A future all-sky satellite mission with sensitivity $\delta I_\nu = 0.1$ Jy sr$^{-1}$
and spectral resolution $\delta\nu=15\ghz$ can detect the recombination lines at 5$\sigma$ 
for frequency coverage $30\leq\nu\leq 600\ghz$. If higher frequency channels are included 
($30\leq\nu\leq 3000\ghz$), a sensitivity of $\delta I_\nu \simeq 0.25$ Jy sr$^{-1}$ is 
needed for a 5$\sigma$ detection.
An experiment with milli-Jansky sensitivity in frequency channels $\delta\nu=1\ghz$ like 
{\small MILLIMETRON} may measure the recombination lines with a SNR of $\simeq 2500$.
\end{itemize}

A few more comments are in order. Firstly, the CIB fluctuations are produced by fluctuations 
in the distribution of high-redshift galaxies and, therefore, should correlate strongly with 
the fluctuations measured in large scale structure (LSS) surveys provided the latter are 
deep enough to resolve stellar masses of order $M_\star\gtrsim 10^{10-11} \hmsun$. Therefore, 
it may be possible to remove the CIB substantially if we overlap with a deep LSS survey. 

Secondly, a major issue for the detection of the recombination signals, and any of the 
primordial distortions really, will be the calibration. To separate different frequency 
dependent components, a calibration down to the level of the sensitivity is required. In this 
way, different channels can be compared and the frequency-dependent signals can be separated. 
For the recombination signal, it may be enough to achieve sufficient channel cross-calibration, 
since in contrast to the primordial $\mu$ and $y$-distortion the signal is quite variable. 
This issue will have to be addressed in the future.

Finally, we highlight that a detection of the recombination signal also guarantees a detection 
of the low-redshift $y$-distortion and the small-scale dissipation signals. These two signals 
are expected in the standard cosmological model and allow us to address interesting questions 
about the reionization and structure formation process, as well as the early Universe and 
inflation physics. Even with a sensitive low resolution CMB spectrometer it may furthermore be 
possible to transfer some of the absolute calibration to an independent high-resolution CMB imager. 
This could further open a possibility to extract the line-scattering signals from the dark ages 
and the recombination era \citep{Kaustuv2004, Jose2005, Lewis2013Ray}, which in terms of 
sensitivity are also within reach.

In summary, detection of the  hydrogen and helium lines from recombination of the early universe is a 
hugely challenging but highly rewarding goal for the future of CMB astronomy. This field, largely 
neglected for three decades, is ripe for exploitation.
An all-sky experiment is required to measure the recombination lines. This can best be done from space, 
or possibly  via long duration balloon flights. Likely rewards would include the first spectroscopic 
study of the very early universe and  the first measurement of the primordial helium abundance, as well 
as unsurpassed probes of new physics and astrophysics in a an entirely new window on the earliest epochs 
that we can ever directly access by astronomical probes.

\small
\section*{Acknowledgments}
VD would like to thank Marco Tucci for discussions, and acknowledges support by the Swiss National 
Science Foundation. JS acknowledges   discussions with S. Colafrancesco. JC is supported by the Royal 
Society as a Royal Society University Research Fellow at the University of Cambridge, U.K. 
Part of the research described in this paper was carried out at the Jet Propulsion Laboratory, California 
Institute of Technology, under a contract with the National Aeronautics and Space Administration. 

\small
\bibliographystyle{mn2e}
\bibliography{references}

\label{lastpage}

\end{document}